\documentclass[iop]{emulateapj} %[apj]{emulateapj}
\usepackage{amsmath,amssymb,bm}
\usepackage{graphicx}
 \usepackage{times}
 \usepackage{mathptmx}  % use Times fonts if available on your TeX system
\usepackage{comment}
\usepackage{url}
\usepackage{color}
\usepackage[colorlinks, citecolor=blue]{hyperref}

\newcommand{\Bep}{B_\mathrm{ep}}
\newcommand{\Bet}{B_\mathrm{et}}
\newcommand{\VA}{V_\mathrm{A}}
\newcommand{\tauA}{\tau_\mathrm{A}}

\begin{document}

\title{Helical Kink Instability in a Confined Solar Eruption}

\author{Alshaimaa Hassanin\altaffilmark{1,2} and Bernhard Kliem\altaffilmark{1}}
\email{hassanin@uni-potsdam.de}

\altaffiltext{1}{Institute of Physics and Astronomy, University of Potsdam, 
                 14476 Potsdam, Germany}
\altaffiltext{2}{Department of Astronomy, Space Science \& Meteorology, 
                 Faculty of Science, University of Cairo, Cairo, Egypt}

\shorttitle{Confined solar eruption}
\shortauthors{Hassanin \& Kliem}

% \slugcomment{ms\_paper1\_rev; Aug 12, 2016} %\today} 
\journalinfo{Manuscript ms\_paper1, revised 2016 Aug 12} % 1st revision
\submitted{Received 2016 June 2; accepted 2016 August 22}

\begin{abstract} \noindent 
A model for strongly writhing confined solar eruptions suggests an origin in the helical kink instability of a coronal flux rope which remains stable against the torus instability. This model is tested against the well observed filament eruption on 2002 May~27 in a parametric MHD simulation study which comprises all phases of the event. Good agreement with the essential observed properties is obtained. These include the confinement, terminal height, writhing, distortion, and dissolution of the filament, and the flare loops. The agreement is robust against variations in a representative range of parameter space. Careful comparisons with the observation data constrain the ratio of the external toroidal and poloidal field components to $\Bet/\Bep\approx1$ and the initial flux rope twist to $\Phi\approx4\pi$. Different from ejective eruptions, two distinct phases of strong magnetic reconnection can occur. First, the erupting flux is cut by reconnection with overlying flux in the helical current sheet formed by the instability. If the resulting flux bundles are linked as a consequence of the erupting rope's strong writhing, they subsequently reconnect in the vertical current sheet between them. This reforms the overlying flux and a far less twisted flux rope, offering a pathway to homologous eruptions. 
\end{abstract}

\keywords{Instabilities -- magnetohydrodynamics (MHD) -- Sun: corona -- 
          Sun: coronal mass ejections (CMEs) -- Sun: flares -- 
          Sun: magnetic fields}

% \clearpage ~ \clearpage 

\section{Introduction}
\label{s:intro}

Erupting magnetic flux on the Sun often remains confined in the corona without evolving into a coronal mass ejection (CME). The rise of the flux then halts and any embedded filament or prominence material slides back to the bottom of the corona along the magnetic field lines \cite[e.g.,][]{HJi&al2003}. Confined and ejective eruptions begin similarly \citep{Moore&al2001}; both forms are usually associated with a flare. Their initially accelerating rise indicates the onset of an instability. Confined (or ``failed'') eruptions present an important testbed for theories of solar eruptions. Understanding what prevents an evolving eruption from becoming ejective is also relevant for the study of the space weather and its terrestrial effects \cite[e.g.,][]{Gosling1993, Webb&al2000}. 

In eruption models based on ideal MHD instability \citep{vanTend&Kuperus1978}, confinement results when the condition for the torus instability (in general terms: a sufficiently rapid decrease of the coronal field with height) is not met at or above the eruption site. This is possible if the eruption is caused by the onset of the helical kink instability in the stability domain of the torus instability \citep{Torok&Kliem2005}. If the helical kink saturates before the rising flux reaches the height range where the torus instability can act, then the eruption remains confined; otherwise a CME results. Another possibility arises if the coronal field is structured such that the condition for the torus instability is fulfilled in two separate height ranges enclosing a stable height range, as has been found, e.g., by \citet{YGuo&al2010} and \citet{ZXue&al2016a}. If slowly rising current-carrying flux reaches the lower unstable range, it erupts due to the torus instability, but is halted in the stable height range. 

Another model for confined eruptions 
suggests that the reconnection of two magnetic loops may yield two stable new loops, while producing a flare due to the release of magnetic energy during the reconnection \citep{Nishio&al1997, Hanaoka1997}. Observational support for this model was based on low-resolution data, which did not clearly reveal the nature of the interacting loop-shaped structures \cite[e.g.,][]{Green&al2002}. Coronal magnetic loops are no longer considered to contain sufficient free magnetic energy to power an eruption, rather the much larger amount of flux and free energy typically contained in a filament channel appears to be required. The suggested scenario indeed occurred in an event that showed the reconnection between two filaments \citep{Torok&al2011, Joshi&al2014a}. A CME was associated, but must have originated in the perturbed flux overlying the filaments which remained in place. \citet{YJiang&al2013, YJiang&al2014} reported partly similar (more complex) failed filament eruptions. Overall, events of this category are very rare, however. 

The confinement of eruptions may also be related to the existence of a coronal magnetic null point which spans a dome-shaped magnetic fan surface above the eruption site. In the case of an eruption, the footprint of the fan surface yields a circular flare ribbon \citep{Masson&al2009}. 
This configuration has been found to be associated with both confined eruptions \cite[e.g.,][]{HWang&CLiu2012, NDeng&al2013, Vemareddy&Wiegelmann2014, Kumar&al2015} and ejective ones \cite[e.g.,][]{CLiu&al2015, Joshi&al2015, Kumar&al2016}. Therefore, the fan surface of a coronal magnetic null does not appear to be the primary factor deciding the ejective vs. confined nature of eruptions launched under it. In fact, reconnection at the null may facilitate the removal of overlying, stabilizing flux \citep[e.g.,][]{XSun&al2013, CJiang&al2014}. 

Observational studies of confined eruptions have become quite frequent with the recent advances of observing capabilities \citep{YLiu&al2009, YShen&al2011, YShen&al2012, Netzel&al2012, Kuridze&al2013, HChen&al2013, HQSong&al2014, SYang&al2014, Kumar&Cho2014, RLiu&al2014, Joshi&al2014b, Kushwaha&al2014, Kushwaha&al2015, XCheng&al2015, TLi&JZhang2015, ZXue&al2016b}. Statistical studies of their association with source region structure and flare magnitude were also performed \citep{YWang&JZhang2007, XCheng&al2011}. Despite the large variety of these events, several trends are apparent. Strong overlying flux is typically observed, and for several events it was demonstrated to possess a height profile that prevents the torus instability. Naturally, this favors the central part of active regions above their periphery. Indications of the helical kink and the dissolution of embedded filaments are often found. 
Confinement also shows an association with the magnitude of energy release by the eruption, with only a minor fraction of the B-class flares but nearly all $>$X1-class flares being accompanied by CMEs. The underlying cause-effect relationship, i.e., whether the confinement limits the flare magnitude or an insufficient flare energy release prevents the eruption from developing into a CME, is not yet clear. However, since there are about three orders of magnitude between the weakest CME-associated flares (in the low-B-class range) and the strongest confined flare, the flare magnitude cannot be decisive by itself but must be considered in the context of the source region's magnetic structure, particularly the properties of the overlying flux. 

Occasionally, even eruptions producing X3 flares can remain confined, as in the exceptional active region (AR)~12192. Insufficient magnetic shear, twist or helicity in the AR's core field and overlying flux preventing the torus instability have been suggested to be the possible causes of confinement in this region \citep{Thalmann&al2015, XSun&al2015, HChen&al2015, JJing&al2015, Inoue&al2016, LLiu&al2016}. Since a strong eruption (the X3 flare) occurred in the first place, confinement by the overlying flux appears to be the most obvious explanation. 

Here we present a simulation study of the confined filament eruption on 2002 May~27, whose detailed and comprehensive observations were analyzed by \citet{HJi&al2003} and \citet{Alexander&al2006}. The initial phase of the eruption, up to the point the rising flux reached its terminal height, was already modeled by \citeauthor{Torok&Kliem2005} (\citeyear{Torok&Kliem2005}, henceforth TK05). Using a flux rope susceptible to the helical kink mode but not to the torus instability as the initial condition in their MHD simulations, the rise profile of the flux rope apex, the rope's developing helical shape, and its distortion during the deceleration showed close agreement with the observations. Thus, the helical kink instability appears to be the prime candidate mechanism for this event. Our simulations substantiate this model for confined eruptions in two ways. First, we extend the computations to model the whole event, and second, a parametric study suggests that the requirement on the initial twist can be relaxed to about $4\pi$, which is closer to twist estimates for other events than the estimate of $\approx5\pi$ by TK05. We also focus on the magnetic reconnection, demonstrating that it occurs in two distinct locations and phases which correspond to the observed brightenings and changes of topology, and consider the fate of the erupting flux, which can reform a (less twisted) flux rope.

\section{Observations}
\label{s:observations}

\begin{figure}[!t]                                                % Fig. 1
 \centering
 \includegraphics[width=.98\linewidth]{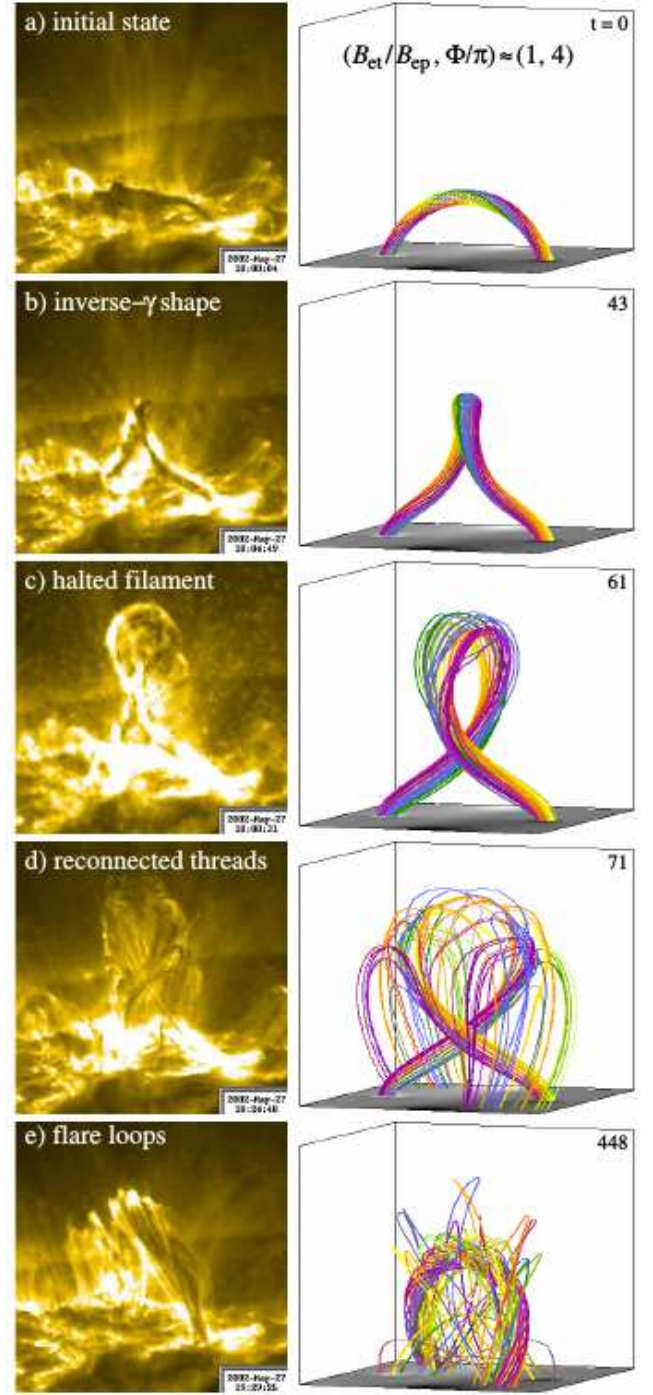}
 \caption[caption]{Characteristic stages of the confined filament eruption in AR~9957 on 2002 May~27 observed by \textsl{TRACE} at 195~{\AA} (\textit{left column}) compared with the best matching Case~1-4 of the parametric simulation study (\textit{right column}). \textsl{TRACE} images are rotated by 90~deg. Magnetic field lines in the volume $5\!\times\!5\!\times\!4$ in the center of the box are shown, and the magnetogram, $B_z(x,y,0)$, is included in gray scale. The first four panels show field lines in the core of the kink-unstable flux rope (with start points on a circle of radius $a/3$ centered at the rope axis) and the final panel shows mostly ambient field lines that were first reconnected with, and subsequently disconnected from, the field lines shown above in the course of the two main reconnection phases. \\\hspace{\textwidth} 
(An animation of the right column is available. 
 An animation of the \textsl{TRACE} images is available at \url{http://trace.lmsal.com/POD/movies/T195_020527_18M2.mov}.)}
\label{f:shape_best}
\end{figure}

The eruption of a filament in AR~9957 near the west limb of the Sun on 2002 May~27 commencing at about 18~UT is a classical case of a confined event. Detailed analyses of the EUV observations by \textsl{TRACE} \citep{Handy&al1999} and X-ray observations by \textsl{RHESSI} \citep{RLin&al2002} were presented in \citet{HJi&al2003} and \citet{Alexander&al2006}, so that here we only give a brief summary of the most relevant observations. The filament was observed by \textsl{TRACE} in the 195~{\AA} band at high resolution (of 0.5~arcsec per pixel) and cadence (of up to 9~s). A strong writhe develops during the rise, resulting, for the given perspective of the observer, in an inverse-gamma shape. The filament reaches the peak projected height of 84.4~Mm at 18:09:31~UT, experiencing a strong distortion of its upper part, which is followed by the dissolution into a heterogeneous group of loop-shaped threads. The threads form new connections and eventually disappear to give way to a set of bright flare loops in the same place, appearing in the 195~{\AA} band from 19:03~UT onward. After saturating the detector for $\gtrsim10$~min, they begin to turn into absorption at 19:17~UT. The \textsl{TRACE} observations of these characteristic phases are shown in Figure~\ref{f:shape_best}. There are no signs of a CME in the data of the LASCO coronagraph \citep{Brueckner&al1995}. 

\begin{figure}[!t]                                                % Fig. 2
 \centering
 \includegraphics[width=.95\linewidth]{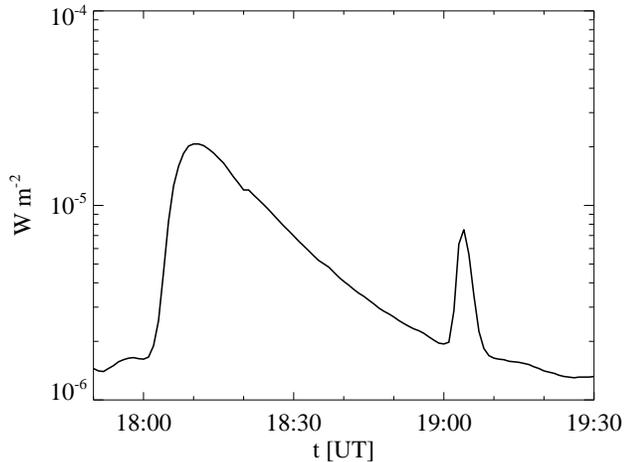}
 \caption{\textsl{GOES} soft X-ray light curve of the M2.0 flare associated with the eruption. An unrelated C7.5 flare originating at the east limb is superimposed after 19~UT.}
 \label{f:GOES}
\end{figure}

\begin{figure}[!t]                                                % Fig. 3
 \centering
 \includegraphics[width=.95\linewidth]{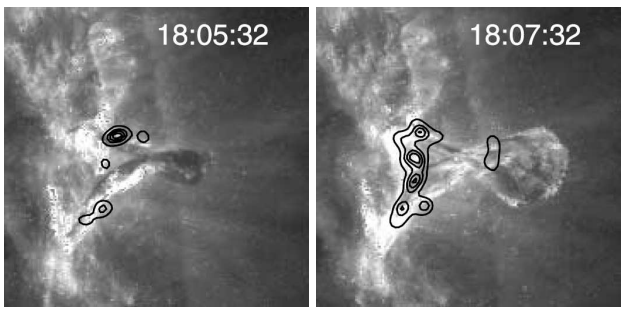}
 \caption{\textsl{TRACE} 195~{\AA} and \textsl{RHESSI} 12--25~keV observations in the rise phase of the confined eruption \cite[(with permission) from][]{Alexander&al2006}.}
 \label{f:TRACE+RHESSI}
\end{figure}

\begin{figure}[!t]                                                % Fig. 4
 \centering
 \includegraphics[width=.95\linewidth]{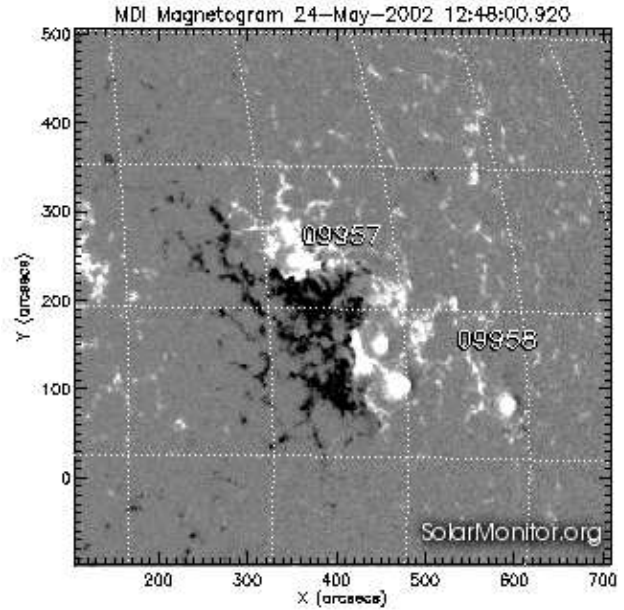}
 \caption{MDI line-of-sight magnetogram of AR~9957 (and the adjacent small AR~9958) on 2002 May~24, 12:48~UT.}
 \label{f:AR9957}
\end{figure}

The accompanying soft X-ray (SXR) flare of \textsl{GOES} class M2.0 commences near 18:01~UT and reaches its peak at 18:10~UT (Figure~\ref{f:GOES}). Hard X-rays (HXRs) observed by \textsl{RHESSI} are emitted from the very beginning of the filament's rise and peak at 18:06:40~UT in the 12--25~keV band. EUV brigthenings in the 195~{\AA} band also commence simultaneously with the filament's rise and intensify from 18:04:30~UT, i.e., at about half of the rise. Initially, the strongest HXR sources and much of the brightest 195~{\AA} emissions are located close to the rising filament, primarily at its top side. At about 18:06~UT the positions of the strongest sources switch to the bottom of the corona along a line connecting the footpoints of the filament (see Figure~\ref{f:TRACE+RHESSI} and the full image sequence in \citealt{Alexander&al2006}). Additionally, a weak HXR source, peaking nearly simultaneously with the 12--25~keV HXRs from the whole AR, occurs near the crossing point of the filament's legs (mostly slightly above the crossing point; also see Figure~\ref{f:TRACE+RHESSI}). This source stimulated the suggestion that the filament legs reconnected with each other in this eruption \citep{Alexander&al2006}. (Similar evolution of the HXR sources was observed in other strongly kinking filament eruptions; see \citealt{RLiu&Alexander2009}.) The band of bright EUV emission near the bottom of the corona (Figure~\ref{f:shape_best}(c)--(d)) does not evolve into a set of flare loops; such loops rather appear at greater heights of $\approx$~35--50~Mm and only late in the event (Figure~\ref{f:shape_best}(e)). 

The details of the photospheric flux distribution as well as the exact path and footpoints of the filament are difficult to discern, due to the proximity to the limb. Figure~\ref{f:AR9957} shows AR~9957 and the adjacent, near unipolar small AR~9958 on May~24, 3.25 days before the eruption. Both ARs are decaying, and the magnetograms from the MDI instrument \citep{Scherrer&al1995} near the time of the eruption indicate that no principal change of their structure occurs in the meantime, i.e., the changes appear to consist of the standard processes of flux dispersal and cancellation. Prior to the eruption, cool filamentary plasma can be seen in the \textsl{TRACE} 195~{\AA} images along the whole polarity inversion line of AR~9957, but only the section at solar-$y\lesssim230$ erupts. The 195~{\AA} flare ribbons and flare loops are also exclusively formed along this section (Figure~\ref{f:shape_best}). 
Therefore, a bipolar source model appears justified. The southern footpoint of the erupting filament section is rooted in negative photospheric polarity, indicating right-handed field.

\section{Numerical model}
\label{s:numerics}

As in TK05 we integrate the compressible ideal MHD equations in the limit of zero plasma beta and gravity,
%=========================================================================
\begin{eqnarray}
\partial_t\rho&=&
                 -\bm{\nabla\cdot}(\rho\,\bm{u})\,,      \label{eq_rho}\\
\rho\,\partial_{t}\bm{u}&=&
      -\rho\,(\,\bm{u\cdot\nabla})\,\bm{u}
      +\bm{J\times B} 
      +\bm{\nabla\cdot\mathsf{T}}\,,                     \label{eq_mot}\\
\partial_{t}\bm{B}&=& 
    \bm{\nabla \times} (\bm{u\times B})\,,               \label{eq_ind}\\
\bm{J}&=&\mu_0^{-1}\,\bm{\nabla\times B}\,,              \label{eq_cur}
\end{eqnarray}
%=========================================================================
using a modified Lax-Wendroff scheme \citep{Torok&Kliem2003}. Here $\bm{\mathsf{T}}$ denotes the viscous stress tensor 
($\mathsf{T}_{ij}=\rho\,\nu\,[\partial u_{i}/\partial x_j+
                            \partial u_{j}/\partial x_i-
                       (2/3)\delta_{ij}\,\bm{\nabla\cdot u}]$)
and $\nu$ is the kinematic viscosity, included to prevent numerical instability. In the modified Lax-Wendroff scheme, the stabilizing, but rather diffusive Lax step is replaced by a small amount of so-called artificial smoothing \citep{Sato&Hayashi1979}, applied to $\bm{u}$ and $\rho$, but not to $\bm{B}$. This smoothing consists in replacing 5 percent of the value of the variable on each grid point by the average of its six neighbors after each iteration step \cite[see, e.g.,][]{Torok&Kliem2003}. Although far less diffusive than the Lax step, which replaces the whole value of the variable, this level of smoothing suffices to prevent numerical instability even when the evolution becomes highly dynamic at small scales after the onset of magnetic reconnection. Reconnection occurs in the simulations due to the intrinsic numerical diffusion of the magnetic field in the second-order finite-difference scheme whenever current sheets steepen sufficiently.

The simplifying assumption $\beta=0$ is very well justified for eruptions that originate in active regions, where $\beta\sim10^{-4}\mbox{--}10^{-2}$ low in the corona \citep{Gary2001}. The neglect of gravity is justified as we do not attempt to model the return of the lifted filament plasma, which begins to slide back to the bottom of the corona along the field lines as soon as the rise phase of the eruption ends.

Similar to TK05, we use the Titov-D\'emoulin (TD) model of a force-free flux rope equilibrium in bipolar ambient field \cite[][hereafter TD99]{Titov&Demoulin1999} as the initial condition. This approximate analytical equilibrium allows us to perform an extended parametric study. For the range of parameters considered here, force-free equilibrium is well approximated. The configuration consists of three elements. A toroidal current channel of major radius $R$ and minor radius $a$, running in the center of a magnetic flux rope, is partially submerged such that the symmetry axis of the torus runs horizontally at depth $d$ under the photosphere, which is represented by the bottom boundary of the computation box. The current in the current channel consists of a toroidal and a poloidal component, namely, the toroidal ring current $I$ along the channel and, perpendicular to that, an azimuthal or poloidal component, so that inside the current channel the field lines of current density and magnetic field have the same helical form, 
as required by a force-free equilibrium. This structure models the filament. A pair of subphotospheric magnetic sources of strength $\pm q$, placed at the symmetry axis at distances $\pm L$ from the torus plane, provides the external poloidal field component, $\Bep$ (occasionally called the ``strapping field''), which enables the equilibrium by balancing the Lorentz self-force of the current channel \citep{Shafranov1966}. This yields two flux concentrations in the photosphere near the positions $x=\pm L$ which resemble the pair of main sunspots in a bipolar active region. Finally, an external toroidal field component, $\Bet$, is introduced by a line current $I_0$ running along the symmetry axis. This component models the shear field typically seen in active regions, in particular, the axial field of the filament channel that hosts the filament. The field by the line current is known to prevent ejective eruptions (\citealt{Roussev&al2003}; TK05; \citealt{Myers&al2015}), except for very small values of $I_0$. This feature is consistent with the present purpose of modeling a confined eruption. 

The initial density is set to $\rho_0(\bm{x})=|\bm{B_0}(\bm{x})|^{3/2}$, also as in TK05. This scaling with the initial field, $\bm{B_0}(\bm{x})$, ensures a slow decrease of the Alfv\'en velocity, $\VA$, with distance from the flux concentrations, as is typical in the corona. In particular, the resulting height profile $\VA(z)$ closely matches the coronal height profile inferred in \citet{Vrsnak&al2002} from solar radio bursts emitted by coronal shocks (see their result for $\beta\ll1$, perpendicular shock propagation, and five-fold Saito density). The system is at rest initially, except for a small initial velocity perturbation applied at the flux rope apex, which is detailed below.
The initial apex height of the toroidal axis of the current channel and flux rope, $h_0=R-d$, the initial field strength $B_0$, density $\rho_0$, and Alfv\'en velocity $V_\mathrm{A0}$ at that position, and the resulting Alfv\'en time, $\tauA=h_0/V_\mathrm{A0}$, are used to normalize the variables. 

The computations are performed on a stretched Cartesian grid which resolves the volume $|x|\le5$, $|y|\le5$, $0\le z\le10$ with $141\times319\times505$ grid points such that the resolution is high ($\Delta=0.02$) and nearly uniform in the central part ($|x|\la0.9$, $|y|\la2.8$) and gradually degrades toward the side boundaries but not with height. Closed boundaries are implemented throughout by setting $\bm{u=0}$ in the boundary layer. In the bottom boundary, this keeps the normal component of the magnetogram invariant.

\section{Parametric Study}
\label{s:parametric}

\begin{table*} %[b] %[h]                                                  % Tab. 1
\caption{Overview of the parametric study (see text for the further parameters of the equilibria). 
The dimensional values of $a$, $L$, $\tau_\mathrm{A}$, and $V_\mathrm{A0}$ are obtained from the scaling in Section~\ref{sss:scaling}. 
The ratio $\Bet/\Bep$ is given at the magnetic axis of the flux rope. 
}
% \vspace{12pt} 
\centering 
\begin{tabular}{ccccccccc}
\hline
 Case  & $I_0/I_{00}$ 
               & $a$ [Mm] 
                     & $L$ [Mm] 
                          & $\Bet/\Bep$ 
                                 & $\Phi$ 
                                            & $n(h_0)$ 
                                                   & $\tau_\mathrm{A}$ [s] 
                                                          & $V_\mathrm{A0}$ [km\,s$^{-1}$]\\
\hline
 1.4-4 &  1.0  & 8.7 & 43 & 1.37 & $4.0\pi$ & 0.86 &  7.1 & 3400  \\ % fila2.All/pert_1/fila2_phi4pi.All/L_105_n_48_4pi
 1-5   &  0.7  & 6.9 & 45 & 1.03 & $5.0\pi$ & 0.75 & 14.6 & 1500  \\ % fila2I_07.All/pert_1/fila2I_phi5pi/L_120_n_48_500
 1-4   &  0.7  & 9.0 & 57 & 1.15 & $4.0\pi$ & 0.62 &  9.4 & 2600  \\ % fila2I_07.All/pert_1/fila2I_phi4pi/L_140/n_48_500/
 1-3.5 &  0.7  & 9.7 & 54 & 1.15 & $3.5\pi$ & 0.62 &  4.5 & 5200  \\ % fila2I_07.All/pert_4/fila2I_phi3.5pi/L140
 0.8-4 &  0.4  & 9.1 & 66 & 0.80 & $4.0\pi$ & 0.48 & 10.2 & 2300  \\ % fila2I_04.All/pert_1/fila2Ic_phi4pi/L_3.4/n_48_r0.149  % nopert ex., is unstable, kinks upward
\hline
\end{tabular}
\label{t:1}
\end{table*}

A twisted flux rope in force-free equilibrium is susceptible to two relevant modes of configuration change, both of which are referred to as a form of kink instability in the plasma physical literature. One of these is the torus instability. In the case of an arched flux rope, this primarily leads to an expansion of the rope in the major toroidal direction. Thus, it is a form of the lateral kink. The torus instability is primarily controlled by the height profile, or in full toroidal symmetry by the major radial profile, of the external poloidal field, parameterized by the decay index $n=-\mathrm{d}\,\log\,\Bep(z)/\mathrm{d}\,\log\,z$. The critical (threshold) decay index lies between unity and about two (depending on various conditions), with a canonical value of $n_\mathrm{cr}\approx3/2$ \citep{Kliem&Torok2006, Demoulin&Aulanier2010, Zuccarello&al2015}. 

The other relevant mode is the helical kink instability with azimuthal mode number $m=1$, in the astrophysical literature often simply referred to as the kink instability. This instability is primarily controlled by the flux rope twist, here expressed as twist angle $\Phi(r)=lB_\phi/(rB_\zeta)$, where $(r,\phi,\zeta)$ are local cylindrical coordinates referring to the magnetic axis of the rope and $l$ is the rope length. For a line-tied flux rope, relevant in the solar atmosphere, the critical (threshold) value for the case of a uniformly twisted rope is $\Phi_\mathrm{cr}=2.49\pi$ \citep{Hood&Priest1981}. Non-uniformity of the radial twist profile appears to raise the critical average twist (see \citealt{Torok&al2004} and the references therein). The modes with higher $m$ in a force-free flux rope also appear to have a higher threshold \citep{vanderLinden&Hood1999}. 

If an external toroidal field component is present, the changing flux rope must bend and compress this flux. Hence, $\Bet$ acts to stabilize the rope and raises the cited thresholds, which were obtained for $\Bet=0$. Consequently, for appropriate values of $\Bep(z)$, $\Bet$, and $\Phi$, the flux rope can be in a kink-unstable but torus-stable equilibrium (where ``kink'' refers to the helical kink mode). This is the rationale of the model for confined eruptions by TK05. 

Testing this model by comparison with a specific event should preferably include a parametric study which considers the variation of $\Bep$, $\Bet$, and $\Phi$ within plausible ranges. A first such study was performed by TK05, who varied $\Bet$ and $\Phi$. By changing the line current $I_0$, i.e. the strength of $\Bet$, they matched the terminal height of the kink-unstable flux rope to the observed value. For the resulting $\Bet$ the threshold value of the twist, averaged over the cross section of the current channel, is about $3.5\pi$ \cite[for the averaging expression see][]{Torok&al2004}. The closest agreement with the observed shape of the erupting filament was obtained for $\Phi=5\pi$ (from varying $\Phi$ in steps of $\pi$). 

In the present paper we extend the parametric study of TK05 by varying the strength of $\Bet$ (i.e., the line current $I_0$), the average twist $\Phi$ (i.e., the minor radius $a$ for each selected $I_0$), and the spatial scale of $\Bep$ (i.e., the source distance $L$). The force balance in conjunction with the normalization of field strength fixes the strength of $\Bep$ at the position of the current channel for given $R$ and $a$, but the spatial scale of $\Bep$, i.e., its decay index, can still be varied (by varying $L$). Our strategy here is to consider $I_0$ and $\Phi$ as free parameters and change $L$, for each considered $I_0$ and $\Phi$, until the ratio of terminal height, $h_\infty$, and distance between the footpoints, $D_\mathrm{f}=2(R^2-d^2)^{1/2}$, matches the observation, $h_\infty/D_\mathrm{f}\approx1.1$. For our chosen value of $d/R$ (see below), this is equivalent to $h_\infty\approx3.6h_0$. Since $\Bep$ controls the torus instability, which is suggested to decide the confined vs.\ ejective nature of an eruption \cite[TK05;][]{YGuo&al2010}, this strategy appears to be the natural choice for the present task. The optimum $I_0$ and $\Phi$ are then determined by comparison with the further observed properties of the eruption. This yields a better founded estimate of the magnitude of $\Bet$ in the modeled event. In case a smaller $\Bet$ results, also a smaller $\Phi$ may match the data. Recent estimates of twist in erupting solar flux found values up to about $4\pi$ \citep{YGuo&al2013, TLi&JZhang2015, RLiu&al2016}---not much but systematically smaller than the estimate in TK05, motivating us to address this option.

We consider line currents in the range $(0.4\mbox{--}1)I_{00}$, where $I_{00}=4.5\times10^{12}$~A is the value used in TK05. The average twist is varied in the range $3.5\pi\le\Phi\le5\pi$ in increments of $\pi/2$. For the geometry of the flux rope and the dimensional value of the sources of $\Bep$ we start from the values chosen in TD99 and TK05 to represent an average solar active region, i.e., $R=110$~Mm, $d=50$~Mm, and $q=10^{14}$~Tm$^2$. Given the fact that only the basic geometrical aspects of the complex field in AR~9957 can be approximated by the TD model, we leave the laborious option of additionally varying the geometry, i.e., increasing $d/R$ to obtain a flatter flux rope, for future work (see Section~\ref{ss:shape} for a brief discussion). 
These choices of $R$ and $d$ yield an initial flux rope height of 60~Mm and, with the requirement $h_\infty\approx3.6h_0$, a terminal height of $\approx216$~Mm. In order to match the observed terminal height of 84.4~Mm, the dimensional length values must be multiplied by $\approx84.4/216=0.39$, resulting in $R\approx42.9$~Mm, $d\approx19.5$~Mm, and $h_0\approx23.4$~Mm. In Section~\ref{sss:scaling} we will scale each of our simulation runs individually to the observation data. This fixes the dimensional values of the length parameters individually with a slight scatter of less than $\pm5$~percent about these nominal values. 
It should also be noted that the normalization by $B_0$ changes the values of $q$, $I$, and $I_0$ by the same factor. Table~\ref{t:1} provides an overview of the \mbox{$I_0$-$\Phi$} combinations analyzed in detail in the following. These exemplify the behavior of the system in the considered range of parameters. Parameter values obtained by scaling the simulation results to the observations, i.e., $L$, $\tau_\mathrm{A}$, and $V_\mathrm{A0}$, are also given. 

All runs are started with a small velocity perturbation in order to exclude a downward kinking of the flux rope apex, which is preferred above the upward kinking by the TD equilibrium for some parameter combinations \cite[see][]{Torok&al2004}. We prescribe a small upward-directed velocity in a small sphere of radius $a$ centered at the apex of the flux rope for a few Alfv\'en times. The perturbation velocity is linearly ramped up from zero at $t=0$ to a maximum and then simply switched off. For all runs with $\Phi\ge4\pi$ the helical kink mode is weakly or moderately unstable and develops out of the numerical noise if the equations are simply integrated in time. The only function of the initial perturbation for these runs is to guarantee that the kink is directed upward. Uniform values of, respectively, $0.02~V_\mathrm{A0}$ and $2\tauA$ are chosen for the peak upward perturbation velocity and ramp-up time in these runs. For $\Phi=3.5\pi$ the helical kink mode turns out to be stable for $I_0=0.7I_{00}$. Here the initial perturbation has the primary function to push the flux rope into the kink-unstable domain. The strength of the perturbation must be adjusted when $L$ is varied for the matching of the observed terminal height; larger $L$ values (i.e., stronger $\Bep$ above the flux rope) require somewhat stronger perturbations. For the optimum $L$ in this case, the perturbation is ramped up to $0.04~V_\mathrm{A0}$ in $4\tauA$.

\section{Results}
\label{s:results}

\subsection{Confined Eruption Comprising two Phases of Reconnection}
\label{ss:reconnection}

All simulations performed in this study yield a dynamic behavior in basic agreement with the key aspects of the observations throughout the event: 
(1) the writhing of the erupting flux according to the $m=1$ helical kink mode which yields the observed shape, 
(2) the confinement at the observed height, 
(3) the subsequent dissolution of the erupting flux by reconnection with the overlying flux, and 
(4) the formation of the final loop arcade by a second phase of reconnection. 
A continuation of the simulation in TK05 (which is not presented here) yields the same basic agreement. In Figure~\ref{f:shape_best} we show this for the Case~1-4 which matches the \textsl{TRACE} observations best. Quantitative information in the following description also refers to this case. 

Following the prescribed initial perturbation, the eruption starts with an exponentially growing rise (linear phase of the helical kink instability) up to $t\approx40$. The characteristic helical shape develops clearly and in agreement with the observed inverse-gamma shape. This represents the conversion of flux rope twist into writhe of the rope's axis \citep{Torok&al2010}. The corresponding rotation of the rope axis about the vertical reaches $\approx60$~deg. Both current sheets (see below) form in this phase, beginning already at $t\approx2$.

\begin{figure}[!t]                                                % Fig. 5
 \centering
 \includegraphics[width=.95\linewidth]{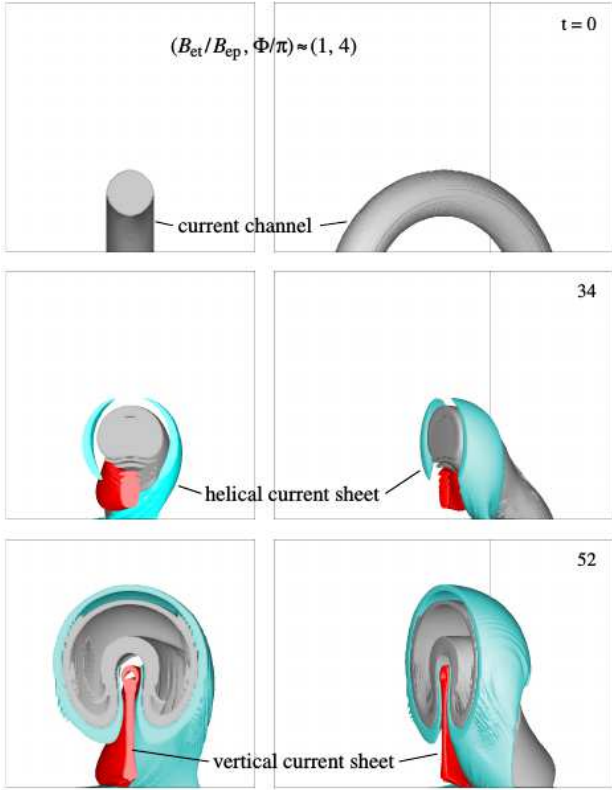}
 \caption{Isosurfaces of current density, $|\bm{J}|$, in Case~1-4, showing the current channel in the center of the TD flux rope (\textit{gray}), the helical current sheet (\textit{cyan}), and the vertical current sheet (\textit{red}) in the volume $4\!\times\!4\!\times\!4$, each in a front view and a side view rotated by 60~deg. An isosurface level of, respectively, 6 and 4 percent of $\max(|\bm{J}|)$ is chosen for $t=34$ and 52. With the exception of the side view at $t=0$, all isosurfaces are restricted to the volume $\{y\ge0\}$, to display the structure in the center of the system.} 
\label{f:iso_J}
\end{figure}

Subsequently, the instability saturates, and the flux rope reaches its terminal height of $\approx3.6h_0$ and maximum apex rotation of 120~deg during $t=60\mbox{--}65$. The saturation occurs by the changing balance between the weakening tension and hoop forces in the twisted flux rope and the increasing back reaction from the pile-up of ambient flux, which steepens the well known helical current sheet in their interface \cite[see Figure~\ref{f:iso_J} and, e.g.,][]{Gerrard&al2001, Kliem&al2010, Kliem&al2014}. The helical current sheet reaches a higher current density than the kinking current channel. The axial direction of the current in the helical sheet is opposite to that in the current channel, so that they repel each other. The rise of the flux rope also produces the vertical current sheet below the rope (Figure~\ref{f:iso_J}), which is known to be the prominent current sheet and a key element in ejective events \cite[e.g.,][]{JLin&Forbes2000}. The current in the vertical sheet attracts the rising current channel. 

\begin{figure}[!t]                                                % Fig. 6
 \centering
 \includegraphics[width=.95\linewidth]{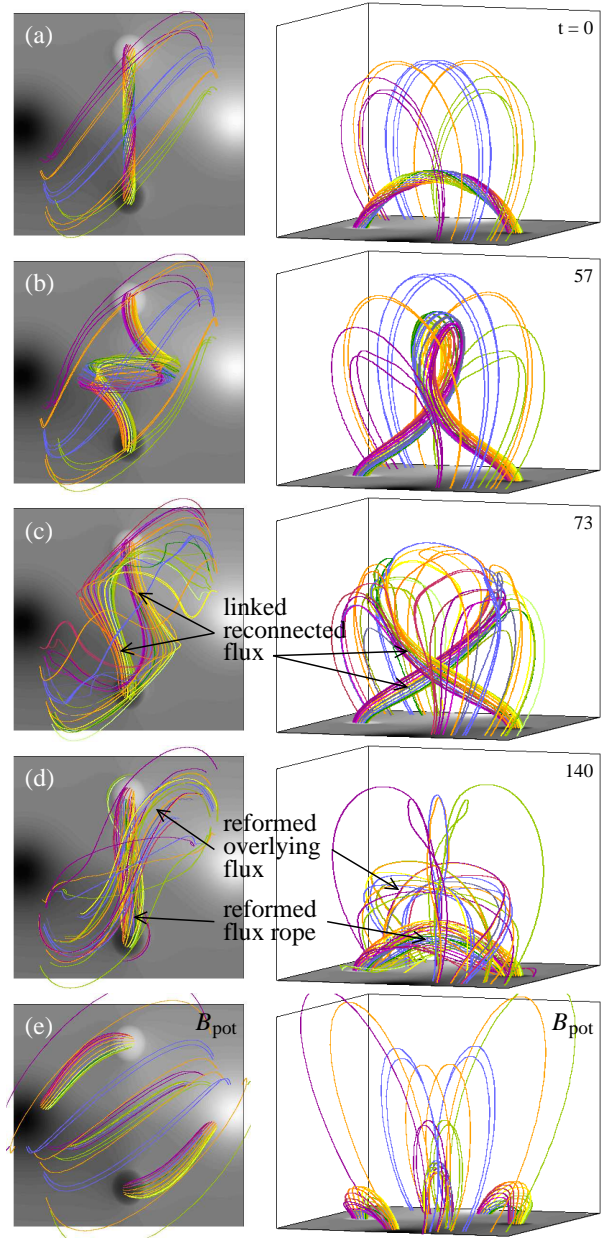}
 \caption{Overview of reconnection and necessity of the second reconnection phase in the considered model: 
 (a) TD equilibrium for Case~1-4, 
 (b) ideal phase of the helical kink instability (already during saturation), 
 (c) two hook-shaped, linked flux bundles after the first reconnection, 
 (d) reformed flux rope and overlying arcade resulting from the second phase of reconnection, and 
 (e) the corresponding potential field. 
 All field lines are traced from start points in the same fluid elements (mostly fixed in the bottom plane; the others moving with the top part of the unstable flux rope).}
\label{f:hooks}
\end{figure}

\begin{figure*}[!t]                                                % Fig. 7
 \centering
 \includegraphics[width=.81\textwidth]{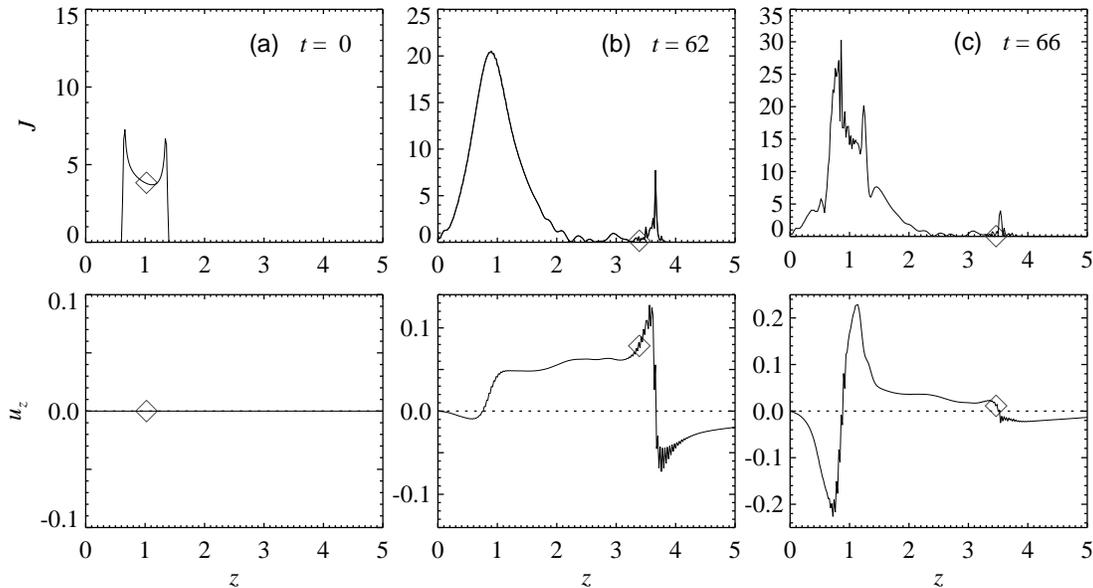}
 \caption{Profiles of current density $|\bm{J}|$ and velocity $u_z$ at the $z$ axis at characteristic times in Case~1-4, showing the current sheets and reconnection flows. The diamond marks the position of the fluid element initially at the apex of the flux rope's magnetic axis. (\textit{a}) initial configuration, (\textit{b}) time of peak reconnection rate in the helical current sheet, (\textit{c}) time of peak reconnection rate in the vertical current sheet.}
\label{f:z_prof}
\end{figure*}

Reconnection plays an important role in the saturation of the instability (Figures~\ref{f:shape_best}, \ref{f:hooks}, and \ref{f:z_prof}). It commences in both current sheets at $t\approx50$, but initially intensifies more prominently in the helical sheet. We refer to this process as the \emph{first reconnection phase}. In all runs we have performed, the flux rope reconnects with overlying ambient flux at various places in this current sheet and eventually is completely cut (at $t\approx75$ in Case~1-4). The reconnection proceeds mostly at two symmetrical X-lines at the sides of the rising flux rope in its upper part (Figures~\ref{f:shape_best}(d) and \ref{f:hooks}(c)), but at times also just above the apex. The latter can be seen at its peak time from the $z$ axis profiles in Figure~\ref{f:z_prof}(b), which show the steepened helical current sheet at $z=3.6$ and its reconnection inflows. The complete reconnection of the flux rope eliminates the driving forces of the rise. 

The field lines reconnected in the helical current sheet are shaped like two linked hooks (Figure~\ref{f:hooks}(c)). The linking is a consequence of the strong writhing (apex rotation); this will be discussed in Section~\ref{ss:reform} below. One side of each hook consists of a leg of the original flux rope, and the other side consists of originally overlying flux. Flux in the leg rotated backward in the view of Figure~\ref{f:shape_best} now finds its other footpoint in front of the other leg which is rotated forward, and the situation is line-symmetrical with respect to the $z$ axis for the other hook. Exactly this topology is shown by several filament threads in the \textsl{TRACE} observations, confirming the reconnection with overlying flux. Moreover, both the range of apex heights and the somewhat irregular arrangement of the new field lines, distributed around the center of the eruption, correspond well to the \textsl{TRACE} data. The new field lines show two directions of motion, toward the center of the system ($z$ axis), which we address below, and downward. As the reconnection progresses, the outer, already reconnected parts of the erupted flux slide down on the surface of the still loop-shaped inner part, distorting the inner flux by dragging it out sideways and downward. This is very similar to the motion of the threads in the upper part of the erupted filament and the resulting distorted appearance. 

It is worth emphasizing again that the reconnection of the flux rope with \emph{overlying} ambient flux progressively \emph{decreases} the flux content of the rope, up to its full destruction. The consequence of this reconnection is opposite to the classical ``flare reconnection'' of ambient flux in the vertical current sheet \emph{under} a rising flux rope, which \emph{increases} the flux in the rope \cite[e.g.,][]{JLin&Forbes2000, Qiu&al2007, Vrsnak2008, Janvier&al2014}. 

The further evolution of the two hook-shaped, linked flux bundles is shown in Figure~\ref{f:hooks}(d). Since the axial currents in the flux bundles have the same main (toroidal) direction and the Lorentz force of bent flux points to the center of curvature, the bundles attract and approach each other, steepening the vertical current sheet in the center of the system, where reconnection of ambient flux is ongoing from $t\approx50$. This leads to a \emph{second phase of strong reconnection} when the legs of the original flux rope come into contact. The upward and downward reconnection outflows from the vertical current sheet strongly amplify during $t\approx63\mbox{--}100$, with the peak values of current density and outflow velocity occurring around $t=66$; see Figure~\ref{f:z_prof}(c). For the parameters of Case 1-4 this reconnection proceeds between the original flux rope legs. Such reconnection is hardly ever observed in ejective eruptions, where, according to the standard picture, purely ambient flux reconnects in the vertical current sheet (see \citealt{Karlicky&Kliem2010} and \citealt{Kliem&al2010} for a possible exception). Since the attractive force of the linked flux bundles continues to act until they are fully reconnected, the necessary outcome is a reformed flux rope low in the volume. This rope is largely complete by $t\approx115$, somewhat less coherent, and considerably less twisted than the original rope. The other two halves of the flux bundles, which are now reconnected, are both rooted in ambient photospheric flux. They straighten out as they are released from the vertical current sheet with the upward reconnection outflow, restoring the arch-shaped overlying flux at heights somewhat below the terminal height of the erupted filament. This corresponds well to the considerable starting height and the late appearance of the observed flare loops. 

Subsequently, reconnection in the vertical current sheet continues at a much lower, gradually decreasing rate. The outflow velocities soon fall into the 0.005--0.01 range, a factor $\sim20$ below their peak values, and stay in this range for a long time. The reconnection is then no longer driven by the erupted flux rope and the current sheet continuously shortens. As a result, the overlying flux gradually adjusts its shape in a few $10^2$ Alfv\'en times (compare Figures~\ref{f:hooks}(d) and \ref{f:shape_best}(e)). 

From a more general perspective, the second reconnection can also be understood as a necessary element in approaching a state of lower energy. Figure~\ref{f:hooks}(e) shows the magnetic connections between the four flux concentrations of the TD magnetogram in the corresponding potential field. The connections are flipped from the ones in the initial TD equilibrium (Figure~\ref{f:hooks}(a)). The writhing of the unstable flux rope turns the legs into the opposite direction, however. In order to approach the potential field, the legs must turn back and reconnect a second time if they are linked after the first reconnection. We will consider this process and its variants for other parameters in more detail in Section~\ref{ss:reform}.

\subsection{Supporting EUV and HXR Observations}
\label{ss:supporting}

\begin{figure}[!t]                                                % Fig. 8
 \centering
 \includegraphics[width=.95\linewidth]{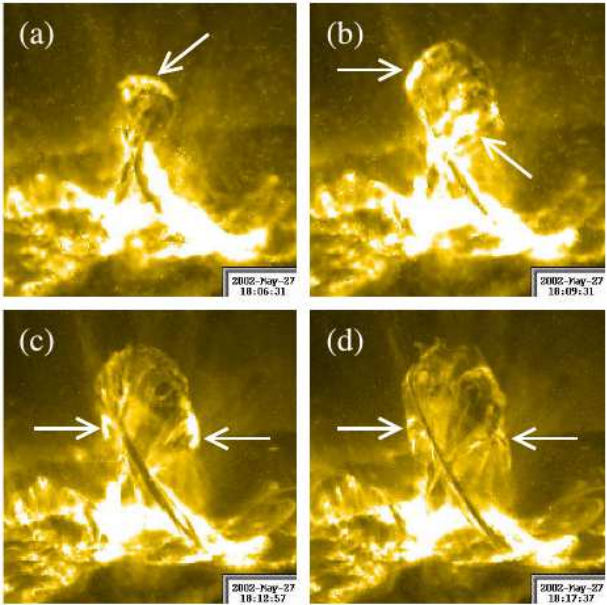}
 \caption{\textsl{TRACE} 195~{\AA} images indicating that reconnection with overlying flux proceeds at various, changing locations, which are marked with arrows in (a)--(c). Reconnected filament threads appear at the position of such brightenings (d).}
\label{f:rec_TRACE}
\end{figure}

In addition to the basic agreement between observation and simulation found above, further specific aspects in the observation data compare favorably with the simulations. Most of these are found to support the model quite strongly, and the other ones appear at least consistent with the model. Figure~\ref{f:rec_TRACE} shows EUV snapshots that correspond to the first reconnection phase in the helical current sheet. The brightenings at various, changing locations indicate the occurrence of reconnection with overlying flux at locations similar to those in the simulations. The two brightenings marked in panel~(c) are particularly persistent. Several reconnected filament threads appear at these locations (panel~(d)), in agreement with the preferential reconnection at two X-lines at similar positions in the simulations. An indication of reconnection at a single X-line at the top of the structure is seen at the time of panel~(a). 

High current densities and the onset of reconnection in the helical current sheet are consistent with the observation of EUV brightenings and HXR sources near the surface and primarily on the top side of the rising filament (see Figure~\ref{f:shape_best}(b)--(c), Figure~\ref{f:TRACE+RHESSI} at 18:05:32~UT, \citealt{HJi&al2003}, and \citealt{Alexander&al2006}). 

The transition of the dominant reconnection in the simulations from the helical to the vertical current sheet corresponds to the transition of the main HXR and EUV sources from the legs of the rising filament to the bottom of the corona under the filament in the course of the rise (around 18:06~UT; see Figure~\ref{f:TRACE+RHESSI} and \citealt{Alexander&al2006}), although reconnection in the helical current sheet likely continues to contribute to these emissions for some time. 

\begin{figure}[!t]                                                % Fig. 9
 \centering
 \includegraphics[width=.95\linewidth]{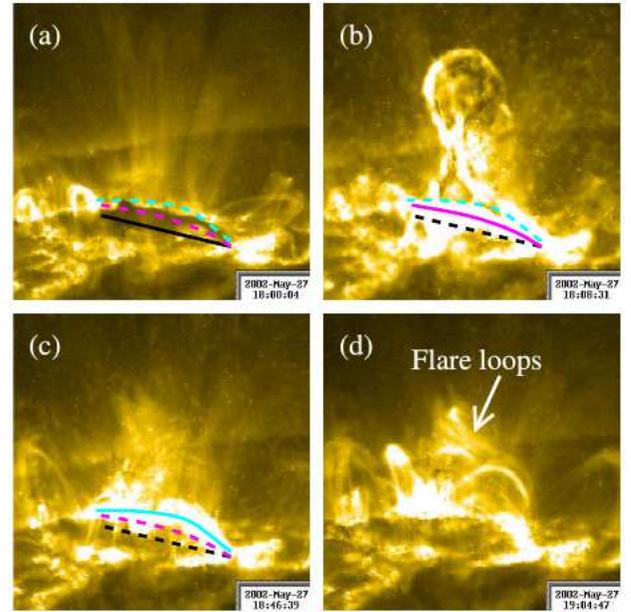}
 \caption{\textsl{TRACE} 195~{\AA} images showing a rise of the brightenings under the erupted filament during the evolution of the disintegrated filament threads and the indication of initial S shape by the flare loops. (a) The visible filament end points prior to eruption are connected by a black line. (Note that the true left (northern) end point of the filament is likely located further westward.) (b) The brightenings before the filament threads reconnect are indicated by the magenta line. They are consistent with a location near the polarity inversion line under the filament. (c) The brightenings just before the reconnecting filament threads disappear, indicated by the cyan line, must have a higher position. (d) The set of dominant flare loops initially indicate an S shape similar to the reformed overlying flux in Figure~\ref{f:hooks}(d).}
\label{f:ref_rope_TRACE}
\end{figure}

Flare loops in ejective eruptions are formed below the reconnecting vertical current sheet. They typically appear in a continuous sequence, growing from rather small heights and footpoint separations and rather high shear to much larger, unsheared arcades of approximately semicircular loops. They begin to form early; even at the temperature of 1.5~MK mainly imaged in the 195~{\AA} band they are often already seen during the impulsive rise phase. There is hardly ever any sign of twist in them. The flare loops in the considered event appear at a considerable height and delay and show only a very minor subsequent rise. This is consistent with a formation above the reconnecting vertical current sheet in the second reconnection phase seen in the simulations. They show a change of shape from moderately S shaped to approximately semicircular (seen nearly edge on), which also corresponds to the simulation (see the initial shape in Figures~\ref{f:ref_rope_TRACE}(d) and \ref{f:hooks}(d) and the final shape in Figure~\ref{f:shape_best}(e)). Indications of twist become quite clear when the flare loops have cooled sufficiently (Figure~\ref{f:shape_best}(e)). This very unusual property is a natural outcome of the reformation process of overlying flux in our simulations because part of the original flux rope joins the reformed overlying flux (see Figure~\ref{f:hooks}(c) and also Section~\ref{ss:reform}). 

The \textsl{TRACE} images also provide indications of reconnected flux low in the corona. The EUV brightenings under the erupted filament show an ascent during the time the disintegrated filament threads evolve and the second reconnection is ongoing; see Figure~\ref{f:ref_rope_TRACE}. Whereas the HXR footpoint sources in Figure~\ref{f:TRACE+RHESSI} (right panel) and the corresponding EUV brightenings in Figures~\ref{f:TRACE+RHESSI} and \ref{f:ref_rope_TRACE}(b) are consistent with a location at the bottom of the corona along the polarity inversion line under the filament, the EUV brightenings in Figure~\ref{f:ref_rope_TRACE}(c) are not. There is no reason to expect that the latter are displaced from the polarity inversion line to the degree that the figure would imply if they were located at the bottom of the corona. Hence, they must be elevated, indicating reformed flux that connects the original footpoints of the filament. Moreover, at times when the \textsl{TRACE} images are somewhat less saturated, these EUV brightenings resemble a collection of very low-lying loops. They remain bright throughout the phase reconnected filament threads are seen in the images ($\sim$~18:10--50~UT, with the images during $\sim$~18:50--19:03~UT being strongly distorted by particle hits on the detector in the Earth's radiation belts). Essentially, this is the decay phase of the SXR flare (Figure~\ref{f:GOES}). The long decay of the SXR light curve clearly indicates ongoing reconnection. 

The set of dominant EUV flare loops become visible after 19:03~UT, i.e., when the reconnection indicated by the SXRs and by the low-lying bright EUV structures significantly decreases. This is consistent with the picture obtained from the more recent multi-band EUV observations of flares with the \textsl{Solar Dynamics Observatory}, which show that flare plasmas are often heated to $T\ge10$~MK before they cool to show flare loops at 1.5~MK in the 193--195~{\AA} band. The low densities at the considerable height of the flare loops in the considered event imply long cooling timescales, making it plausible that the loops appear only when the reconnection decreases significantly. In contrast, the cooling is far more efficient at the higher densities of the low corona, where the bright 195~{\AA} structures in Figure~\ref{f:ref_rope_TRACE}(b) and (c) appear throughout the second reconnection phase. 

The second phase of reconnection indicated by these observations lasts much longer than the second phase of fast reconnection in the simulation. This is likely to result from the more irregular 3D arrangement of the reconnected filament threads (Figures~\ref{f:shape_best}(d) and \ref{f:ref_rope_TRACE}(c)) compared to the hook-shaped reconnected flux bundles in the simulation (Figure~\ref{f:hooks}(c)). 

Some filamentary material reappears along the polarity inversion line under the erupted filament immediately after the event (Figure~\ref{f:shape_best}(e)). This is consistent with the reformation of a flux rope in the simulation, although the new structure is far less coherent than the original filament.

\subsection{Matching of Shape}
\label{ss:shape}

Proceeding to more quantitative comparisons between observation and simulation, we attempt to estimate the external toroidal (shear) field, $\Bet \propto I_0$, and twist, $\Phi$, in the source region from the best match. First, the shapes of the structures formed by the eruption are compared with the corresponding shapes in the simulation. Four characteristic shapes are identified in the \textsl{TRACE} data as displayed in Figure~\ref{f:shape_best}(b)--(e): 
(1) the inverse-gamma shape of the erupted filament which is formed by the fully developed helical kink instability; 
(2) the distorted upper section of the filament at the moment the terminal height is reached; 
(3) the disintegrated and reconnected filament threads resulting from the first reconnection in the helical current sheet; and 
(4) the flare loops resulting from the second reconnection in the vertical current sheet. 

The reconnected filament threads are characterized by the first \textsl{TRACE} image that clearly shows two sets of reconnected threads with a leg in front of the erupted structure (18:24:48~UT). These new connections are traced by cool filament material draining down to the flare ribbon on the front side of the eruption site. (The extent down to the flare ribbon for these and further reconnected threads can be seen most clearly from the moving filament material in the animation of the \textsl{TRACE} data.) The selected time lies well within the interval the reconnected threads are seen or indicated in the data ($\sim$~18:12--18:50~UT). They appear with apex points lying at the side of the erupted filament which still contains threads that extend up to the terminal height of the eruption and are most likely not yet reconnected. The apex points of the reconnected threads lie about halfway between the crossing point of the filament legs prior to their reconnection and the terminal height. The flare loops are characterized by the image at 19:29:25~UT which shows their final shape and the indications of twist most clearly. 

Correspondingly, the field line plots to be compared with the reconnected filament threads are selected at a time a significant fraction of the field lines has reconnected in the helical current sheet and their new apex points lie about halfway between the crossing point of the flux rope legs and the terminal height of the eruption. This also clearly shows the location of their new footpoints. For the comparison with the flare loops we select a snapshot that clearly shows an arch-shaped collection of field lines in the reformed overlying flux near the position of the observed flare loops and no significant further change. For most of our cases, this is reached quite long after the reformation of the flux rope, because the overlying flux passes through many small adjustments of its shape under the influence of ongoing relatively slow reconnection in the vertical current sheet (see the animation accompanying Figure~\ref{f:shape_best}). 

The field lines at $t=0$ and for the first three characteristic shapes are all started from the same points in the bottom plane such that they show the flux of the rope within $r=a/3$. A choice of this magnitude is not only suggested by the thickness of the erupting filament in the first half of its rise, when it is not yet disintegrated and its thickness appears clearly, but also by the properties of a force-free flux rope. The condition of force-freeness couples the axial and radial length scales of the rope. For example, in the uniformly twisted Gold-Hoyle flux rope, the radial length scale equals the pitch of the field lines. Thin flux ropes of moderate twist are possible if the twist profile peaks strongly at the axis of the rope \cite[e.g.,][]{RLiu&al2016}, but for filaments in decaying regions like AR~9957, the current density and twist likely peak at the surface of the flux rope \cite[e.g.,][]{Bobra&al2008}. This is similar to the TD equilibrium \cite[see Figure~\ref{f:z_prof}(a) and][]{Torok&al2004}. Therefore, a weakly twisted flux rope of $\Phi\approx(3\mbox{--}5)\pi$ in AR~9957 should be considerably thicker than the observed filament (if the minor radius of the TD rope in our best matching Case~1-4 were reduced to $a/3$, the average twist would rise to $16\pi$). 

Moreover, the \textsl{TRACE} images give the impression that the filament threads run near the magnetic axis of the erupting structure, i.e., not only in the bottom part of a flux rope, as is often assumed in the literature. During about 18:12:30--18:26~UT the threads in the front leg are nearly straight and at times appear to wind about each other. Both properties are expected for flux near the axis of a flux rope, but not near its outer parts. The very good geometrical correspondence between the displayed field lines near the axis of the TD rope and the observed 
shapes of the filament supports this view. 

\begin{figure*}[!t]                                                % Fig. 10
 \centering
 \includegraphics[width=.98\linewidth]{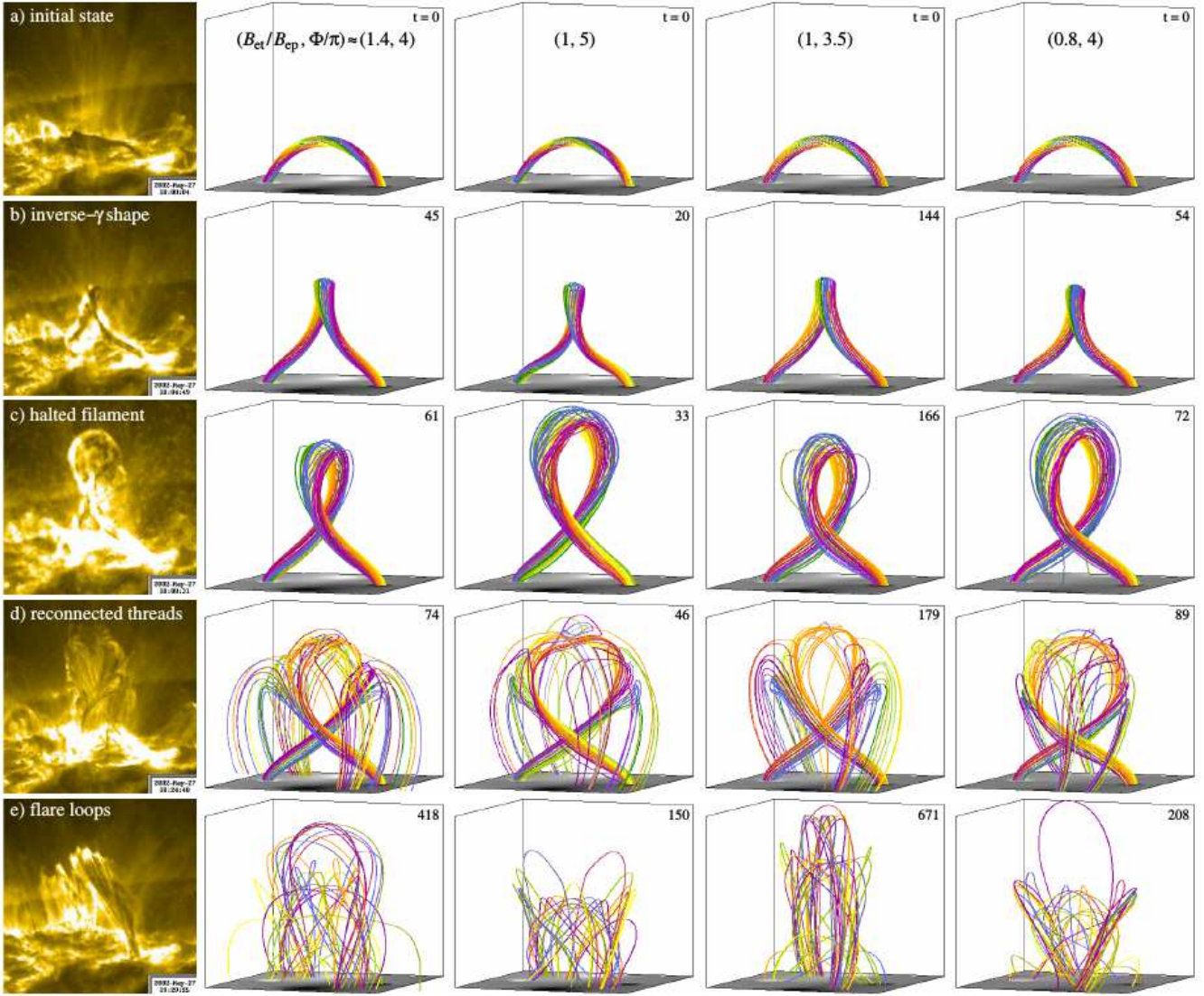}
 \caption{Characteristic stages of the eruption in \textsl{TRACE} 195~{\AA} observations (\textit{left column}), compared with cases~1.4-4, 1-5, 1-3.5, and 0.8-4 (\textit{columns~2--5}). The format of the field line plots is identical to Figure~\ref{f:shape_best}.}
\label{f:shape_all}
% \vspace{4cm} 
\end{figure*}

The five cases are compared with the \textsl{TRACE} data in Figures~\ref{f:shape_best} and \ref{f:shape_all}, with the best matching Case~1-4 shown in Figure~\ref{f:shape_best}. Overall, all five cases show an excellent match of the inverse-gamma shape, a good to very good match of the shape when the maximum height is reached, and a relatively good match of the filament thread shapes during the phase of reconnection in the helical current sheet. Strong differences among the cases are only found for the prominent flare loops in the final stage of the event, which are relatively well matched only by Case~1-4 and roughly matched by three other cases. 

\emph{Inverse-gamma shape.} 
Since the inverse-gamma shape develops very well in all our cases, we pay attention to the fact that the southern filament leg shows a clear S shape. This feature was used in TK05 to fix the best matching flux rope twist to be $\Phi=5\pi$ (the legs were rather straight for $4\pi$ and too helical for $6\pi$ and higher). As expected, Figure~\ref{f:shape_all} shows that the S shape of the flux rope legs is more pronounced for higher $\Phi$ and lower $\Bet$. The Cases~1-5, 1-4, and 0.8-4 yield an excellent match. 

\emph{Distortion of the halted filament.} 
The distortion of the filament's upper section tends to develop most clearly for high $\Bet$ when the overlying flux resists the eruption strongly. We find it most pronounced in Cases~1-4 and 1-5 and in the case (1.4-5) studied in TK05. For the other three cases it develops as well, but only immediately after the end of the rise. We have also measured the height of the leg crossing point at the end of the rise relative to the terminal height. Here the \textsl{TRACE} data (0.35) are closely matched by Cases~1-3.5 (0.33) and 1-4 (0.32), quite well matched by Cases~1-5 (0.30) and 1.4-4 (0.40), and still reasonably matched by Case~0.8-4 (0.27). 

\emph{Reconnected filament threads.} 
All simulation runs match the observation data in the key feature: the new loops are formed in the distorted upper section of the erupted flux by reconnection with overlying flux. Differences are mainly found in the footpoint locations of the new loops. These clearly depend on the main direction of the ambient flux, i.e., on the magnitude of $\Bet$, since $\Bep$ is fixed at the initial flux rope position by the equilibrium condition and thus does not vary strongly among our cases in the volume immediately above the flux rope. The dominance of $\Bet$ in the overlying flux can clearly be seen in Case~1.4-4, where the outer legs of the new loops are strongly aligned in the toroidal ($y$) direction. As $\Bet$ decreases by the factors 0.7 and 0.4, this alignment changes progressively toward the $x$ direction of $\Bep$. The latter cases are in somewhat better agreement with the thread shapes observed by \textsl{TRACE}. 

\emph{Flare loops.} 
The main direction of the overlying flux influences the shape of the structures most clearly when this flux is reformed, becoming visible as flare loops. Here our cases differ most obviously. One or two prominent bundles of loops are formed at the end of each simulation run, but their main direction approximates the observed one preferably in the cases with $\Bet/\Bep\approx1$. The initial flux rope twist influences the resulting main direction of the flare loops as well, because it determines the writhing of the flux rope, and thus influences the steepening of the helical current sheet and location of the first reconnection. A moderate twist, i.e. Case~1-4 in Figure~\ref{f:shape_best}, yields by far the best match. The ratio of projected height and footpoint distance is $\approx1.6$ for the observed flare loops and $\sim1$, $\sim1$, $\approx1.6$, $\sim4$, and $\gg1$ for our cases (in the order they are listed in Table~\ref{t:1}) from the field line plots in Figures~\ref{f:shape_best}(e) and \ref{f:shape_all}(e). 

A further geometrical property of interest is how the values of $L$ obtained from matching the terminal height of the eruption (see Table~\ref{t:1}) compare to the distance of the polarities in the source region. We roughly estimated the latter from the magnetogram on May~24 when the AR was located at about 25~deg west (Figure~\ref{f:AR9957}). The center of gravity of the flux distribution then had a distance of about 40~Mm from the unstable section of the inversion line for both polarities. Here the nearly unipolar neighboring AR~9958 was included because it certainly contributed to the flux passing over the filament. The value may have increased somewhat by the ongoing dispersal of the flux in the subsequent 3.25~days up to the eruption, but such minor change, which is difficult to estimate, is not important for our rough comparison. The $L$ values obtained from the simulation runs are all 
relatively close to the estimated distance, but slightly larger, by up to a factor 1.6. 
They clearly show the expected increasing trend for decreasing $\Bet$, reflecting the necessity of a more stabilizing $\Bep$. It is not surprising that the 
model yields $L$ values which are too large, because its toroidal geometry 
yields an initial flux rope apex height above the apex height of the observed flat filament for our chosen ratio $d/R$. 
To prevent the onset of the torus instability, $L$ must therefore be larger than observed. The initial height, $h_0\approx23.4$~Mm, exceeds the observed one (17.4~Mm) by a factor of 1.3 (see also Figure~\ref{f:scaling}), very similar to the excess of the $L$ value for our favored Case~1-4 and to the average excess. Thus, the $L$ values obtained from matching the observed terminal height are fully consistent with the active-region magnetogram, especially for the Cases~1-4 and 1-3.5. 

In summary, the observed shapes of the rising and halted (distorted) filament are well matched in a wide range of the parameters in the TD model. The appearance of the reconnected filament threads is best matched for the reduced line current values ($\Bet/\Bep\lesssim1$). 
The observed flare loops are satisfactorily matched only by Case~1-4.

\subsection{Scaling of the Simulations and Timing}
\label{ss:scaling+timing}

\subsubsection{Scaling to the Observed Rise Profile}
\label{sss:scaling}

\begin{figure*}[!t]                                                % Fig. 11
 \centering
 \includegraphics[width=.65\linewidth]{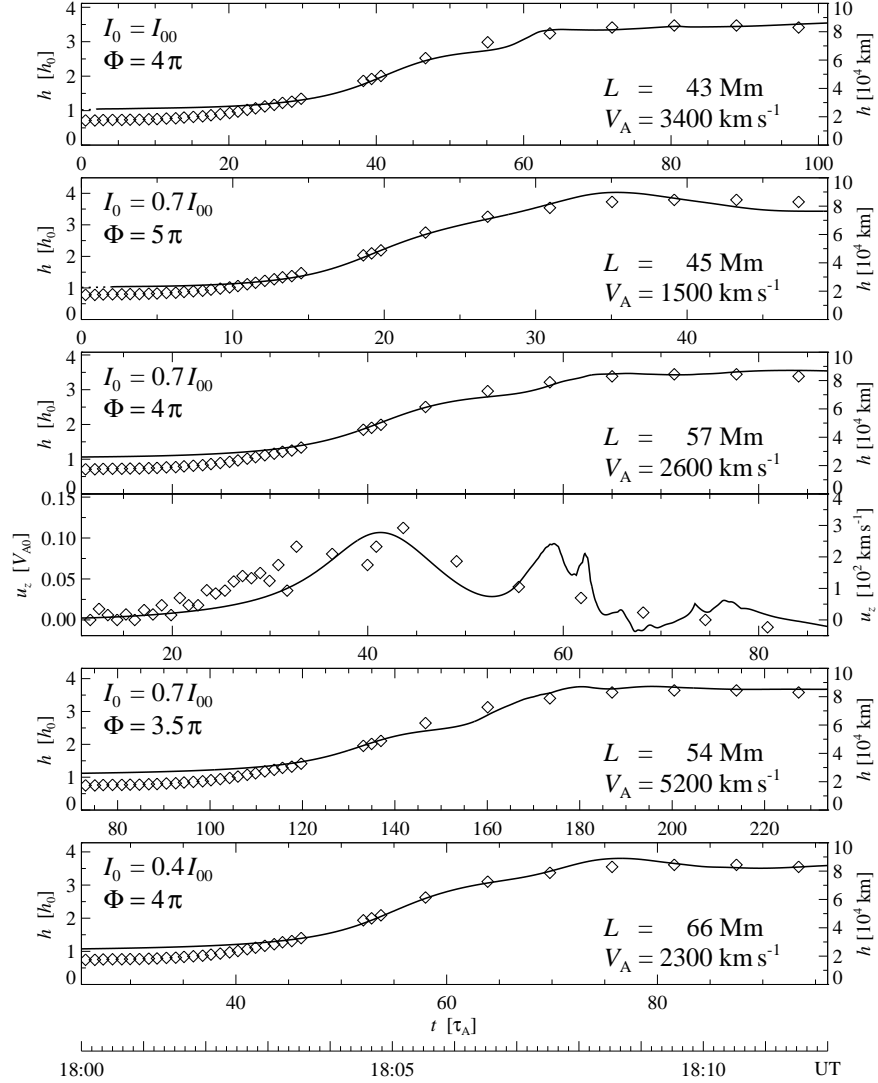}
 \caption{Scaling of the five cases to the observed rise curve of the filament apex. Diamonds show smoothed projected height data from \citet{HJi&al2003} and the derived velocities. 18:08:31~UT is selected as the reference time (see text). The small initial perturbation in the simulations is shown dotted.}
\label{f:scaling}
\end{figure*}

Next we compare the timing of the different phases of the eruption between simulation and observation. This requires a scaling of the time unit in the simulations, $\tau_\mathrm{A}$, to the elapsed time in the observations and an absolute placement of a reference time in the simulations relative to a reference time in the observations. As in TK05, we use the rise profile $h(t)$ and its derivative $u_z(t)$, derived by \citet{HJi&al2003} from the \textsl{TRACE} images, for this purpose. From the simulation runs we extract the rise profile of the fluid element initially at the apex point of the flux rope's magnetic axis, which is a standard output. This height is slightly smaller than the upper edge of the group of field lines plotted in Figures~\ref{f:shape_best} and \ref{f:shape_all} during the rise but reaches the same terminal value. The peak of this rise curve is also slightly delayed from the time of maximum field line height selected in Figures~\ref{f:shape_best}(c) and \ref{f:shape_all}(c), because the fluid element experiences a short phase of additional upward acceleration (discussed below) when the upper edge of the flux rope is halted by the overlying flux. This delay is small, up to 90~s (in Case~1-3.5) and 40~s for our best matching case, and does not influence any of our conclusions. The reference time in the observation data is chosen to be the time the terminal height is essentially reached (18:08:31~UT, with 1~min and only 1.4~Mm of further rise remaining). 

Figure~\ref{f:scaling} shows the results of the scaling. 
For each of our five cases, an optimal match of the observation data for the rise phase and terminal height was sought. This results in an individual scaling of the length unit for each case, which falls in the range $h_0=(22.3\mbox{--}24.5)$~Mm. 
All five cases, as well as the case presented in TK05, reproduce the observed rise profile quite well. They all deviate weakly in the initial height, which is too high in the simulations due to the geometrical restriction of the TD model to a torus. The scaling yields the dimensional values for the distance of the photospheric flux concentrations, $L$, and for the Alfv\'en velocity in the source volume of the eruption, $V_\mathrm{A0}=h_0/\tau_\mathrm{A}$, which are listed in the figure and in Table~\ref{t:1}. 

The estimates of $V_\mathrm{A0}$ depend on the growth rate of the instability through the dimensional value of $\tau_\mathrm{A}$, which increases with increasing growth rate. Thus, $V_\mathrm{A0}$ decreases with increasing twist $\Phi$ and decreasing external toroidal field $\Bet$. The Alfv\'en velocity in the low corona of active regions is not well known. Values in the range $V_\mathrm{A0}\sim1000\mbox{--}2000$~km\,s$^{-1}$ are often quoted. However, values higher by a factor of 10 or even somewhat more appear realistic as well if typical field strengths and densities estimated from radio bursts low in the corona \citep{Dulk&McLean1978} are adopted. Therefore, the estimates of $V_\mathrm{A0}$ must be considered acceptable for all our cases.

\subsubsection{Timing of Instability Onset and Reconnected Structures}
\label{sss:timing}

Since the scaling procedure uses the observed rise profile of the filament apex point, the phase of inverse-gamma shape and the arrival at the terminal height in the simulations are assigned the corresponding observation times. The resulting assignments for the onset time of the instability, formation time of the reconnected hook-shaped loops (reconnected filament threads), and reformation time of the overlying flux (flare loops) in the simulations can be used to check the simulations against the observations. 

\emph{Onset time of instability.} This yields a perfect match for Cases~1.4-4 and 1-5 and a slightly early onset of the instability by $\approx1.5$~min before the first height data point (at 18:00:04~UT) for Case~1-4, by $\approx5.5$~min for Case~1-3.5, and by $\approx4$~min for Case~0.8-4. This comparison is not a sharp discriminator between our cases because the initial development of the filament is very slow, so that the onset time of the rise in the \textsl{TRACE} data cannot be determined sufficiently precisely. A similar judgement results if the typical coupling between CME acceleration and flare X-ray emission \citep{Zhang&Dere2006} is considered. A very close coupling would favor our first three cases. However, \citet{Bein&al2012} found that the CME acceleration onset precedes the onset of the SXR flare emission in about 75~percent of all events with an average difference of 4--5 minutes. The differences obtained from our scalings stay in this range.

\emph{Formation time of reconnected filament threads.} As discussed in Section~\ref{ss:shape}, we use the time the reconnected threads are first seen clearly on the front side of the erupted flux (18:24:48~UT, which is about one third into the time range reconnected threads are visible), and the corresponding time in the simulations is given by the selections for Figures~\ref{f:shape_best}(d) and \ref{f:shape_all}(d). It should here be kept in mind that the latter times refer to the progress of reconnection near the axis of the flux rope ($r\le a/3$), while reconnection commences much earlier at the periphery of the rope ($r\approx a$). The simulation time elapsed from 18:00:04~UT and the ratio of the elapsed times are given in Table~\ref{t:2}. The elapsed times in the simulation are shorter by factors of $\sim2.5$, i.e., the kink-driven reconnection in the helical current sheet proceeds much faster than in reality. We conjecture that the perfect coherence of the TD flux rope is the main reason for this difference. The observed filament is far from being such coherent, as is most clearly seen from the arrangement of the reconnected filament threads which result from the reconnection in the helical current sheet. There is a weak trend for the cases with the higher growth rates to match better. However, the differences between the cases are much smaller than their difference to the observations; hence, they are insufficient for favoring some cases above the others.

\begin{table}[!ht]                                                  % Tab. 2
\caption{Timing of the reconnected filament threads. The observation time is $t_\mathrm{obs}=$18:24:48~UT. The corresponding simulation time elapsed from 18:00:04~UT, $\Delta t_\mathrm{sim}$, dimensionless and scaled, 
and the ratio of elapsed times, $\Delta t_\mathrm{sim}/\Delta t_\mathrm{obs}$, are listed.}
% \vspace{6pt} 
\centering 
\begin{tabular}{ccccccc}
\tableline\tableline
 Case    & $\Delta t_\mathrm{sim}$ [$\tauA$] 
                 & $\Delta t_\mathrm{sim}$ [s] 
                         & $\Delta t_\mathrm{sim}/\Delta t_\mathrm{obs}$\\ 
\tableline
 1.4-4 &   74  &  524  &  0.35 \\
 1-5   &   46  &  673  &  0.45 \\
 1-4   &   61  &  571  &  0.38 \\
 1-3.5 &  106  &  474  &  0.32 \\
 0.8-4 &   64  &  648  &  0.44 \\
\tableline
\end{tabular}
\label{t:2}
\end{table}

\begin{table}[!ht]                                                  % Tab. 3
\caption{Timing of the flare loops relative to 18:00:04~UT. The observation time is $t_\mathrm{obs}=$19:10:41~UT.}
% \vspace{6pt} 
\centering 
\begin{tabular}{ccccccc}
\tableline\tableline
 Case    & $\Delta t_\mathrm{sim}$ [$\tauA$] 
                 & $\Delta t_\mathrm{sim}$ [s] 
                          & $\Delta t_\mathrm{sim}/\Delta t_\mathrm{obs}$\\ 
\tableline
 1.4-4 &  412  &  2930  &  0.69 \\
 1-5   &  150  &  2190  &  0.52 \\
 1-4   &  438  &  4130  &  0.97 \\
 1-3.5 &  599  &  2670  &  0.63 \\
 0.8-4 &  183  &  1860  &  0.44 \\
\tableline
\end{tabular}
\label{t:3}
\end{table}

\begin{table}[!ht]                                                  % Tab. 4
\caption{Timing of the flare loops relative to 18:24:48~UT. The observation time is $t_\mathrm{obs}=$19:10:41~UT.}
% \vspace{6pt} 
\centering 
\begin{tabular}{ccccccc}
\tableline\tableline
 Case    & $\Delta t_\mathrm{sim}$ [$\tauA$] 
                 & $\Delta t_\mathrm{sim}$ [s] 
                          & $\Delta t_\mathrm{sim}/\Delta t_\mathrm{obs}$\\ 
\tableline
 1.4-4 &  338  &  2410  &  0.87 \\
 1-5   &  104  &  1520  &  0.55 \\
 1-4   &  377  &  3560  &  1.29 \\
 1-3.5 &  492  &  2200  &  0.80 \\
 0.8-4 &  119  &  1210  &  0.44 \\
\tableline
\end{tabular}
\label{t:4}
\end{table}

\emph{Formation time of flare loops.} 
We use the time the flare loops are first seen clearly in their whole extent down to the prominent flare ribbons. This is the case at 19:10:41~UT (earlier than the time of the \textsl{TRACE} image in Figures~\ref{f:shape_best}(e) and \ref{f:shape_all}(e), which best displays the internal structure of the flare loop arcade, as explained in Section~\ref{ss:shape}). Subsequently, apart from a very minor, gradual increase in height, there is no significant change in the position and shape of the flare loops. The corresponding times in the simulations are given by the selections for Figures~\ref{f:shape_best}(e) and \ref{f:shape_all}(e). The results from the scaled simulations are compiled in Table~\ref{t:3} in the same format as in Table~\ref{t:2}. Again, the times elapsed in the simulation are shorter, but here the mismatch is significantly smaller. The mismatch tends to be reduced further by referring to the observed formation time of the reconnected filament threads, 18:24:48~UT. This is compiled in Table~\ref{t:4} and yields agreement with the observations to within 30~percent for the Cases~1.4-4, 1-4, and 1-3.5. This latter comparison refers exclusively to the long phase of weakly driven and relatively slow reconnection in the vertical current sheet, which gradually completes and shapes the reformed overlying flux in the simulation (Section~\ref{ss:reconnection}) and is indicated by the gradual decrease of the SXR flux in the observations (Figure~\ref{f:GOES}). 

The preceding phase of strongly driven, fast reconnection in the vertical current sheet ($t\approx63\mbox{--}100$ in Case~1-4 (Section~\ref{ss:reconnection})) roughly coincides with the interval of flux rope reformation in the simulation (up to $t\approx115$ in Case~1-4). Unfortunately, the \textsl{TRACE} observations only provide tentative indications of a reformed flux rope (the brightenings in Figure~\ref{f:ref_rope_TRACE}(b) and (c) and the subsequent reformation of cool filamentary structures in the same place) without any clear hint on its timing. Therefore, the rate of fast reconnection in the vertical current sheet in the simulations cannot be quantitatively compared with the \textsl{TRACE} data.

\subsubsection{Timing Comparison with the Hard X-ray Emissions}
\label{sss:timingHXR}

Further aspects of interest in timing comparisons are the \emph{HXRs from the whole active region}, which at their peak time mostly originate from footpoint sources under the rising filament, and the appearance of a coronal HXR source near the crossing point of the filament legs. Our simulations do not include the acceleration of particles, but a link to the HXR emissions is given by the observation that the time of peak reconnection rate in eruptions is associated with the peak flux of the high-energy emissions \citep{Qiu&al2004}. It is clear that the bulk of the HXR-emitting electrons must originate in one or both reconnecting current sheets (helical or/and vertical) in the system. The 12--25~keV light curves for both source regions begin to rise early in the eruption (before 18:04~UT), peak in the late stage of the filament rise (18:06:40--18:07~UT), and gradually decrease subsequently \citep{Alexander&al2006}. Thus, the onset of reconnection and the time of peak reconnection rate are weakly delayed in the simulations. In Case~1-4 these times ($t\approx50$ and $t\approx62\mbox{--}66$) correspond to $\approx$~18:06~UT and $\approx$~18:08--18:09~UT, respectively. 

Since reconnection depends on both global and local conditions (e.g., on the imposed inflow velocity and on the current sheet thickness), its onset and peak times depend more sensitively on the parameters in the system than the shapes formed by the erupting flux. In particular, the distribution of flux in the photosphere may play a critical role. AR~9957 shows flux concentrations near the polarity inversion line, whereas no such feature exists in the TD equilibrium. These flux concentrations may cause a faster steepening of the vertical current sheet, resulting in earlier onset and peak times of reconnection in this sheet. Similarly, inhomogeneities in the overlying flux, which certainly exist, may cause a stronger and earlier steepening of the helical current sheet in some places as compared to the simulations which use a smooth model field. This effect is indicated by the location of the first HXR sources and EUV brightenings at one position just on top of the northern leg of the rising filament during $\approx$~18:02--18:06~UT (see Figure~\ref{f:TRACE+RHESSI} (left panel) and the animated \textsl{TRACE} images). Additionally, due to the numerical diffusion, the current sheets in the simulation may steepen less readily than the current sheets in the corona. Given these effects, we consider the timing of the fastest reconnection in our simulations to be in basic agreement with the timing of the main HXR emissions and EUV brightenings. 

The \emph{coronal HXR source} near the crossing point of the filament legs starts at a somewhat later time and intensifies more impulsively than the bulk of the HXR emission from the active region, indicating an independent origin. \citet{Alexander&al2006} suggested that the radiating electrons were accelerated by reconnection in the vertical current sheet between interacting filament legs. However, this process likely causes the bulk HXR emissions after their source locations switch from the rising filament legs to the bottom of the corona under the filament at $\approx$~18:06~UT. A further argument against the suggestion comes from the fact that, for most of the time, the coronal HXR source is located somewhat above the crossing point, where neither the reconnection electric field (i.e., the acceleration) nor the ambient plasma density (i.e., the target density for the energetic particles) are expected to be maximal. The indications for particle acceleration by reconnection in the helical current sheet are strongest when the main HXR sources and most prominent EUV brightenings are located just on top of the rising filament legs, i.e., up to $\approx$~18:06~UT, which precedes the peak of the coronal HXR source. Therefore, it is of interest whether there exist further effects that might have caused this source. 

The simulations indeed show a special phenomenon in the right place and at the right time. This is a strong distortion of the inner part of the current channel, which emanates from the upper tip of the vertical current sheet. Figure~\ref{f:iso_J} at $t=52$ shows the situation when this process begins to develop strongly. The vertical current sheet extends into the current channel from below, such that the original circular cross section is strongly distorted. The resulting Lorentz force drives a further strong upward motion from the tip of the current sheet into the volume of the current channel, which is halted by the tension of the overlying flux at the helical current sheet about 10 Alfv\'en times later. This motion is visible in Figure~\ref{f:scaling} as a second peak of the apex velocity in Case~1-4 during $t=52\mbox{--}63$ and in Figure~\ref{f:z_prof}(b) as an upward flow in the range $z\approx2\mbox{--}3$, which is faster than the reconnection outflow in the range $z\approx1\mbox{--}1.5$. The same phenomenon is clearly indicated by the $h(t)$ curves for the other cases (Figure~\ref{f:scaling}), where it also occurs shortly before the terminal height is reached. Layers of alternating toroidal current direction (negative for the current channel and the vertical current sheet, positive for the helical current sheet) are here pushed together very closely and change their local configuration rapidly. The resulting induced electric fields may be responsible for the acceleration of the electrons whose emission is seen as the coronal HXR source.

\subsubsection{Summary of Timing Comparisons}
\label{sss:timing_summary}

To summarize the comparison of the timing, all our simulations can be scaled to the observed rise profile of the eruption, resulting in a good match which is also consistent with known values of the Alfv\'en velocity in active regions. The further possible comparisons refer primarily to the progress of reconnection and yield a heterogeneous picture. Reconnection in the simulations commences somewhat later than in the observed event. However, as long as it is strongly driven by the helical kink, it proceeds considerably (2--3 times) faster. Both differences are likely due to the simplicity of the initial TD equilibrium, whose external field is much smoother and whose flux rope is much more coherent than the corresponding structures in AR~9957. The rather prompt shift of fast, driven reconnection from the helical to the vertical current sheet corresponds to the shift of the strongest HXR sources from the legs of the erupting filament to the bottom of the corona under the filament. The subsequent much slower, undriven reconnection in the vertical current sheet, which reforms and gradually adjusts the overlying flux, proceeds at a rate comparable to the observed one in three of our cases (1.4-4, 1-4, and 1-3.5), whereas it is too fast by a factor $\sim2$ for the other two. In spite of the differences, the reconnection in the simulations appears to be in basic agreement with the observed timing (and source locations) of the HXR emissions and major EUV brightenings in the associated flare. The weak coronal HXR source slightly above the crossing point of the erupted filament's legs may be due to the strong distortion of the erupted flux by the upward-growing vertical current sheet.

\subsection{Flux Rope Reformation}
\label{ss:reform}

\begin{figure}[!t]                                                % Fig. 12
 \centering
 \includegraphics[width=.95\linewidth]{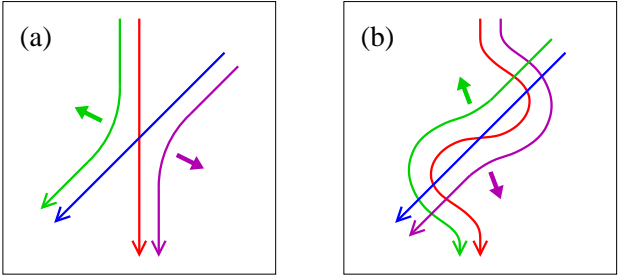}
 \caption{Schematic showing that linked flux bundles (green and magenta in (b)) result from reconnection of a rising flux rope (red) and overlying flux (blue) if and only if the flux rope writhes sufficiently, such that its top part rotates beyond the direction of the overlying flux. The reconnection outflow directions are indicated by filled arrows.} 
\label{f:linking}
\end{figure}

The reformation of the flux rope described in Section~\ref{ss:reconnection} occurs in the whole parameter range studied in this paper; thus, it could be a typical phenomenon in confined eruptions driven by the helical kink instability. As is obvious from Figure~\ref{f:hooks}, the process requires the reconnection of the erupted flux rope with overlying flux in such a way that two linked flux bundles result. This, in turn, requires a sufficient writhing of the erupting flux, such that the flux rope apex rotates (about the vertical) beyond the direction of the immediately overlying flux. It does not require that the reconnection with overlying flux occurs at two X-lines. Figure~\ref{f:linking} illustrates that, after sufficient writhing, linked flux bundles are formed as well if the reconnection proceeds at only one X-line on top of the erupted flux. This can be traced to the fact that the reconnection outflows always align with the antiparallel component of the reconnecting flux. 

For our best matching Case~1-4 with $\Bet\approx\Bep$, the rotation must exceed $\approx\pi/4$, which is a rather modest requirement for the helical kink. Even if the external field is purely poloidal, passing nearly perpendicularly over the polarity inversion line (i.e., if shear field exists only within the current-carrying filament channel), the required rotation of $\gtrsim\pi/2$ can easily be reached by a well developed helical kink \cite[e.g.,][]{Kliem&al2012}. The reformation may be a pathway to a sequence of homologous flares that start with one or several confined eruptions and may include one ejective eruption (CME) at the end \cite[as, e.g., in the events analyzed in][]{YShen&al2011}. More complex sequences are possible if erupting flux is only partly stopped in the corona, which is occasionally observed \citep{RLiu&al2007, Koleva&al2012}. 

Whether the torus instability can lead to the flux rope reformation process found here is not yet clear. Its requirement on the decay index implies a weaker pile-up of ambient flux above the unstable flux rope. Moreover, torus-unstable flux does not writhe considerably by itself. The presence of a sufficiently strong external toroidal field component leads to both stronger writhing \citep{Isenberg&Forbes2007, Kliem&al2012} and flux pile-up, but also has a stabilizing effect. Therefore, further study will be required to determine the potential of the torus instability for the reformation of a flux rope. 

No reformation was found in a previous simulation study of a kink-unstable flux rope which remained confined \citep{Amari&Luciani1999}. Quantitative information about the height profile of the ambient field in that simulation was not given, but, from the geometry shown and from the confinement of the unstable flux rope, one can expect that the ambient field prevented the torus instability from occurring. The flux rope reconnected with the overlying flux; however, different from our simulations, this resulted in two unlinked flux bundles. The reason for this difference lies in the chosen special orientation of the flux rope along the line between the main concentrations of the ambient flux. 

A partial reformation occurs in our model if the legs of the erupted flux rope begin to reconnect with ambient flux before they complete their mutual reconnection, which can be expected if the ambient flux ($\Bet$ or $\Bep$) is sufficiently strong. The effect begins to act when the line current is raised above our reference value, $I_0>I_{00}$. It also occurs when the sources of $\Bep$ (i.e., the sunspots) are closer to the polarity inversion line, leading to a stronger $\Bep$ low in the volume. Both cases lie beyond the range of parametric study in this paper.

Rather small values of twist are found in the reformed flux ropes in the present study. Three of the four effects leading to this result are obvious from Figure~\ref{f:hooks}. The initial twist is first lowered by conversion into writhe. Second, the remaining twist per unit length decreases inversely proportional to the increasing length of the flux rope (in Case~1-4, this is 2.4~times the initial length when the terminal height is reached). Third, only the twist in the two short sections below the crossing point of the legs is transferred to the reformed rope. Additionally, one can expect that the twist tends to equilibrate along each linked hook-shaped flux bundle between the first and second reconnection phases by propagating from the original flux rope leg to the untwisted, originally overlying flux. In Case~1-4 a twist of only $\approx0.7\pi$ results. 

Since the long section of the expanded flux rope above the crossing point becomes part of the reformed overlying flux, that flux necessarily inherits part of the twist in the initial flux rope. This is apparent in the final stage of the runs shown in Figures~\ref{f:shape_best} and \ref{f:shape_all}, most clearly for Cases~1-4 and 1-5.

\section{Conclusions}
\label{s:conclusions}

(1) Using a kink-unstable force-free flux rope in equilibrium as the initial condition, the MHD simulations presented in this paper achieve good agreement with the essential properties of the confined filament eruption in AR~9957 on 2002 May~27, as observed by the \textsl{TRACE} satellite in the EUV. These include 
(i) the confined nature of the eruption, 
(ii) its terminal height, 
(iii) the writhing of the erupting flux according to the $m=1$ helical kink mode which yields the observed inverse-gamma shape, 
(iv) the dissolution of the erupting flux by reconnection with the overlying flux, and 
(v) the formation of the flare loop arcade, which shows indications of twist, by a second phase of reconnection. 
The agreement is obtained in a representative range of parameter space. This robustness supports the model for confined eruptions by TK05, which assumes a kink-unstable and torus-stable flux rope to exist at the onset of the eruption. 

(2) Through quantitative comparisons with the observations, the performed parametric study constrains the ratio of external toroidal (shear) field component and external poloidal (strapping) field component to $\Bet/\Bep\approx1$ and the average twist in the initial flux rope to $\Phi\approx4\pi$, in better agreement with recent twist estimates for other events than the estimate in TK05. 

(3) Different from ejective eruptions (CMEs), the confined eruption triggered by the helical kink is found to comprise two distinct phases of strong reconnection. The first phase occurs in the helical current sheet and destroys the rising flux rope through reconnection with overlying flux (opposite to standard ``flare reconnection'' in ejective events). In the whole range of parameters studied here, this reconnection occurs after a strong writhing of the erupted flux rope, such that the resulting two flux bundles are linked. This, in turn, causes the second phase of strong reconnection between the legs of the original flux rope in the vertical current sheet in the center of the system, which restores the overlying flux. Flare loops are formed in this flux above the vertical current sheet.

(4) The second reconnection also results in the reformation of a flux rope of similar or only moderately reduced flux. Although the twist is strongly reduced (to $\Phi<\pi$ in our simulations), the reformation offers a pathway to homologous eruptions, preferably if the sequence starts with one or several confined events and includes only one CME at the end.

\section{Discussion}
\label{s:discussion}

Of the five cases presented here, one eruption (Case~1-3.5) starts from marginal stability, three from weakly unstable equilibria, and one (Case~1-5) from a moderately unstable equilibrium. No principal differences in the general behavior of the five cases were found and all parametric trends discussed in Section~\ref{s:results} appear to be coherent across them. Therefore, although the magnetic field on the Sun will not jump across the marginal stability line right into the unstable domain of parameter space, it appears justified to study models that employ weakly or even moderately unstable equilibria (as, e.g., in TK05) if done with appropriate caution. 

We have found that the reconnection rate in the numerical model can differ significantly, but not extremely, from the solar one, especially for reconnection strongly driven by an ideal MHD process. Reconnection driven by the helical kink, which pushes the unstable flux rope against the overlying flux, is too fast in the simulations by a factor 2--3. The subsequent undriven reconnection in the vertical current sheet overall proceeds at a rate comparable to (within 30~percent of) the observed rate for three of our five cases, including the best matching case, and is too fast by a factor $\approx2$ for the other cases. The rate of ``flare reconnection'' in a previous modeling of a weak ejective event, estimated to be $\approx1.5$ times lower than on the Sun \citep{Kliem&al2013}, fits into the picture. None of these rates is off by an order of magnitude or more, whereas the numerical diffusion of the code, although intentionally kept at the minimum allowed by the numerical scheme, exceeds the magnetic diffusivity in the corona by many orders of magnitude. It is obvious that the reconnection rate in solar eruptions must primarily be controlled by the large-scale motions of the unstable flux, which act as drivers of the reconnection inflows or outflows. These motions can be modeled with relatively high accuracy by an ideal MHD code, as our comparisons with the observed shapes and rise profile of the considered eruption demonstrate. 

Comparing a parametric simulation study of an analytical model like the present one with the method of data-constrained simulations (where the initial condition typically is a nonlinear force-free field model obtained from the observations by extrapolation or flux rope insertion), one finds specific strengths on either side. The parametric study allows one to disentangle---sometimes discover and often quantify---the relevant effects, isolating the most important ones and their trends. It often also allows matching the observations rather closely, like in the present investigation. Data-constrained modeling can potentially match the observations even better, since the model can account much better for the complexity of the solar field. Less effort for the computations and their analysis may be required. In some cases, new effects can be discovered which are absent in analytical models due to the simplicity of those models. Hence, the modeling strategies 
are complementary.

All eruptions in our parametric simulation study remain confined and show a strong rotation of the flux rope apex about the vertical. The rotation angles in the five cases in Table~\ref{t:1} all reach 120--160~deg. A strong apex rotation (reaching $\sim90$~deg or higher) is also often observed in confined eruptions on the Sun, but is not a typical property of ejective eruptions. Both findings do not support the idea that a considerable apex rotation (as typically resulting from the helical kink instability) facilitates an ejective behavior of eruptions by locally aligning the erupting flux with the overlying flux, such that the erupting flux can pass through the overlying flux without having to open a large part of it \citep{Sturrock&al2001}. The observation and simulation results can be better understood if the cause-effect relationship is reversed. If an eruption is halted, then the magnetic tension of the erupting flux can no longer be relaxed by expansion but only by further writhing, resulting in a tendency for confined eruptions to develop a strong writhing.

\acknowledgments 
We gratefully acknowledge a further analysis of \textsl{RHESSI} data for this manuscript by R.\ Liu and helpful comments by T.\ T{\"o}r{\"o}k, N. Seehafer, R.\ Liu, and the anonymous referee.
We acknowledge the use of data from \textsl{TRACE}, a mission of the Stanford-Lockheed Institute for Space Research, and part of the NASA Small Explorer program, and from \textsl{SoHO}, a joint mission of ESA and NASA. Data in Figure~\ref{f:AR9957} were supplied courtesy of SolarMonitor.org. 
This work was supported by the DAAD and the DFG.

\clearpage 

\bibliographystyle{apj}
\bibliography{paper1}

\end{document}